\newdimen\figsize
\long\def\twoboxesgap#1#2#3{%
  \setlength\figsize{\hsize}%
  \addtolength\figsize{-#3}%
  \divide\figsize by 2
  \makebox{\parbox[t]{\figsize}{#1}%
           \hspace{#3}%
           \parbox[t]{\figsize}{#2}}}

\long\def\twoboxes#1#2{\twoboxesgap{#1}{#2}{2\columnsep}}

\def\qbar{\ensuremath{{\bar q}}}
\def\tbar{\ensuremath{{\bar t}}}
\def\jet{{\rm jet}}
\def\jets{{\rm jets}}
\def\ra{{\rightarrow}}
\newcommand\figref[1]{Fig.~\ref{#1}}
\newcommand\tabref[1]{Table~\ref{#1}}

\def\mathunit#1{\mathop{\hbox{#1}}\mathclose{}\mathord{}}
\newcommand\ipb{\mathunit{pb}^{-1}}                     
\def\tev{\mathunit{TeV}}
\def\gev{\mathunit{GeV}}

\def\pb{\mathunit{pb}}
\def\mch{{m_{H^+}}}
\def\et{\ensuremath{E_T}}
\def\pt{\ensuremath{{p_{\scriptscriptstyle T}}}}
\def\mete{\setbox0=\hbox{$E$}%
          \hbox{$E$\rlap{\kern -0.6em\raise0.08\ht0\hbox{/}}}}
\def\met{\ensuremath{\mete_T}}
\def\dzero{D\O}
\def\htran{\ensuremath{H_T}}
\def\bbar{\ensuremath{{\bar b}}}
\def\gevc{\gev \kern -1.7pt/ \kern -1.7pt c}
\def\gevcc{\gevc^2}
\def\mt{m_t}

\def\ttbar{t\tbar}

\def\BR{{\rm BR}}

\documentclass{hep99}
\usepackage{epsfig}
\usepackage{mymcite}
\graphicspath{{figures/}}
\begin{document}

\title{Recent results from \dzero\ on the top quark}

\author{Scott S Snyder$^1$\\for the \dzero\ Collaboration}

\address{$^1$ Brookhaven National Laboratory, PO Box 5000, Upton NY 11973, USA}

\abstract{We describe three recent results from \dzero\ related to the 
top~quark: a preliminary measurement of the $t\tbar$ spin correlation
in top quark pair production, a search for top quark decays into
charged Higgs bosons, and an improved cross section analysis in the
$t\tbar\ra e\mu$ channel using neural networks.}

\maketitle

\section{Introduction}

Since the observation of the top~quark,
work has continued on further characterizing its
properties~\cite{review}. This note summarizes three recent results
from the \dzero\ experiment at the Fermilab Tevatron, using our full
data sample of $\approx 110\ipb$ from the past collider run:
a measurement of the spin correlation in $t\tbar$
production, a search for top~quark decays to charged Higgs bosons, and 
an improved measurement of the production cross section in the
$t\tbar\ra e\mu$ channel using a neural network analysis.

\section{Top-antitop spin correlation}

At the Tevatron, top quarks are produced mostly in pairs,
via the reaction $q\qbar \ra t\tbar$.  Since the top~quark is a
spin-$1/2$ particle, once a spin quantization axis is chosen, the two
top~quarks in a pair will have either the same or opposite spin
orientations.  In
general, there will be an asymmetry between these two cases.  This
quantity is calculable in the Standard Model (SM); therefore,
any deviation observed from the predicted value would imply
new physics~\mcite{stelzer96,*lee99,*mahlon96,mahlon97}.
We have carried out a preliminary
spin correlation analysis using our data from Run 1~\cite{suyongthesis}.

Uniquely among the quarks, the top quark decays quickly enough that
final state interactions do not perturb its
spin.  Thus, information about the top quark's
initial spin is present in the angular distribution of its decay
products.  This sensitivity is greatest for charged leptons and $d$-type
quarks.
But since it is much easier to identify leptons than $d$-quarks,
we consider only dilepton events, 
in which both top~quarks decay via $t\ra bl\nu$.

\begin{figure}
\centering
\twoboxes
{\epsfig{file=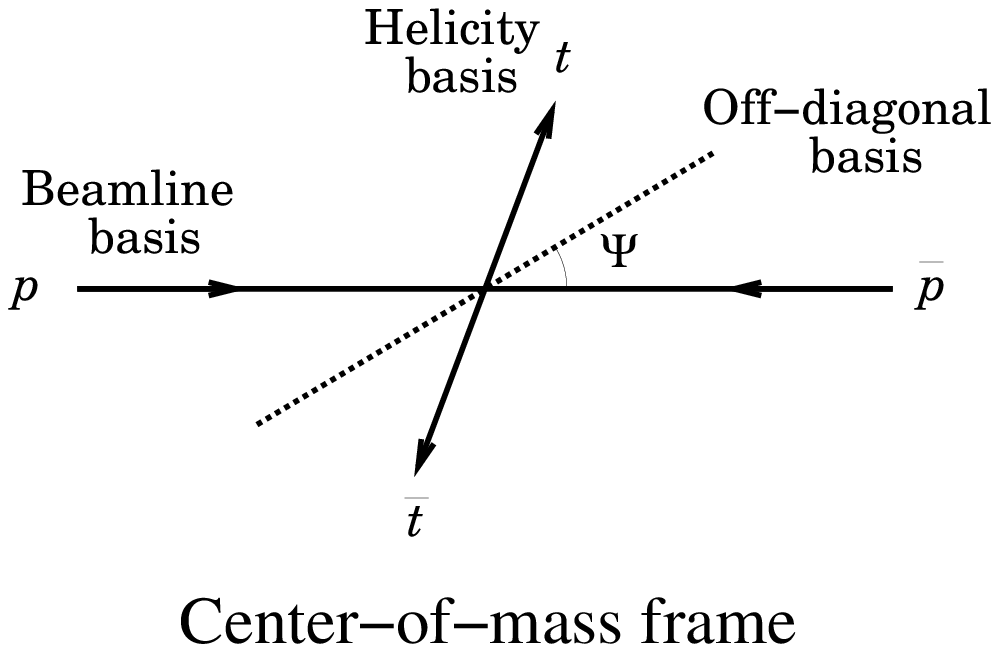, width=\hsize}}
{\epsfig{file=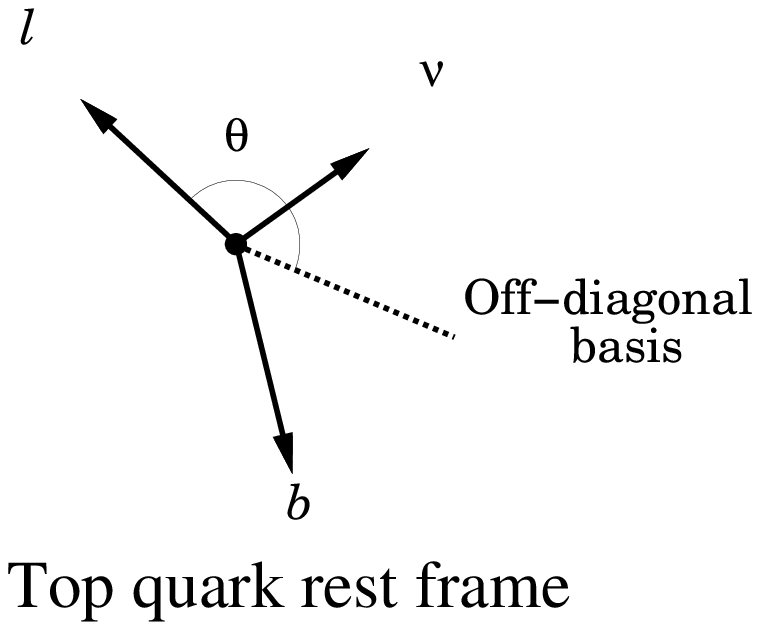, width=\hsize}}
\caption{Definition of the off-diagonal basis and lepton angles $\theta$.}
\label{fg:quanaxis}
\end{figure}

We choose a spin quantization axis known as the ``optimal
off-diagonal basis'' \cite{mahlon97}.
(See \figref{fg:quanaxis}.)  The angle $\Psi$, which depends on the
top quark's momentum and scattering angle, is chosen to give the
largest expected spin correlation.  We define $\theta_{\pm}$ as
the angles between the lepton momenta and the spin axis in the two
respective top~quark rest frames.  Their joint distribution is then
\begin{equation}
  {1\over\sigma} {d^2\sigma \over d(\cos\theta_+) d(\cos\theta_-)} =
  {1 + \kappa\cos\theta_+\cos\theta_- \over 4} .
\end{equation}
The spin correlation information is contained in
$\kappa$;  the SM prediction at
$\sqrt{s} = 1.8\tev$ is $\kappa\approx 0.9$.

With two neutrinos in the final state, the event is
kinematically underconstrained, so we cannot solve directly for
$\theta_{\pm}$.  Rather, we use techniques developed for the
mass measurement~\cite{d0llmassprd} to derive probability
distributions for $\theta_{\pm}$ for each event.

\begin{figure}
\centering
\epsfig{file=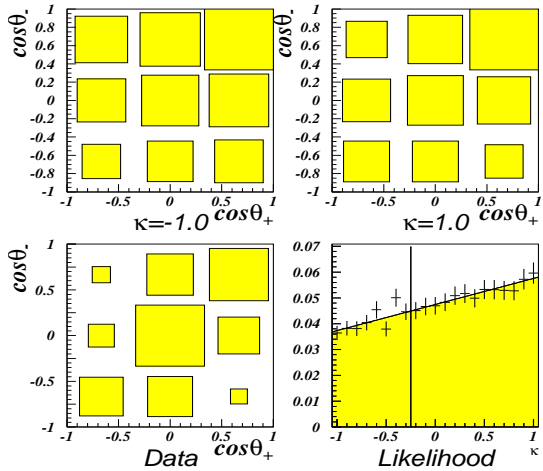, width=\hsize, height=6.5cm}
\caption{Spin correlation results.  The top two
  plots are Monte Carlo simulations for $\kappa=\pm 1$.
  The lower left plot shows the data.  We fit the data to the Monte
  Carlo expectations to derive a likelihood as a function of $\kappa$,
  as plotted in the lower right.}
\label{fg:scresults}
\end{figure}

We have six dilepton candidates, with an
expected background of $1.5\pm 0.3$ events.
For each event, we find the
distribution of solutions in $(\theta_+,\theta_-)$ and bin it
into a 2D histogram.  We then sum over all events.  The
result is shown in \figref{fg:scresults}.  Although the statistics 
from Run~1 are not sufficient to provide a significant
measurement of $\kappa$,we find that
$\kappa > -0.25$, at $68\%$ confidence.


This result will improve greatly during the next
collider run, where we expect about 150
$t\tbar\ra \hbox{dilepton}$ events.
We should be able to
discriminate between $\kappa=0$ and $\kappa=+1$ at least at the
$2.5\sigma$ level, using just the dilepton channel.

\section{Charged Higgs boson search}

The SM contains a single complex Higgs doublet, giving rise to a
single physical Higgs boson, $H^0$.  But one can also consider
extensions to multiple Higgs doublets, as
required by supersymmetry~\cite{hhguide}.
With two Higgs doublets, there are five physical Higgs bosons,
of which two are charged: $H^0$, $h^0$, $A^0$, $H^+$,
and $H^-$.  The electroweak sector is then specified by the
parameters $m_W$, $\mch$, and $\tan\beta$, where $\tan\beta$ is the ratio
of the vacuum expectation values from the two doublets.
If $\mch < m_t - m_b$, then the decay $t\ra H^+ b$ can compete with 
the SM decay $t\ra Wb$.

\begin{figure}
\centering
\epsfig{file=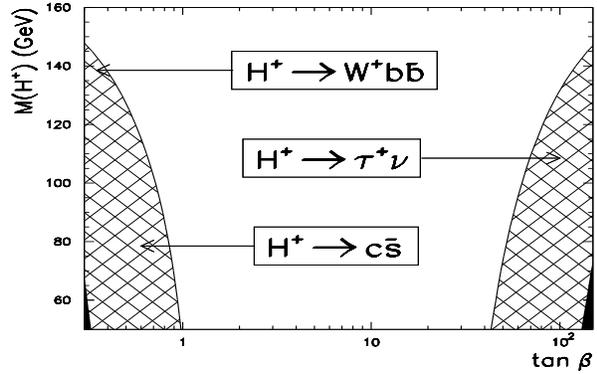, width=\hsize, height=5cm}
\caption{Map of the $(\mch,\tan\beta)$ parameter space; regions with 
  $\BR(t\ra H^+b) > 0.5$ are hatched. The charged Higgs decay modes
  in those regions are also shown.  Regions with
  $\BR(t\ra H^+b) > 0.9$ (dark shaded regions) are not considered.}
\label{fg:chmap}
\end{figure}

\begin{figure}
\centering
\epsfig{file=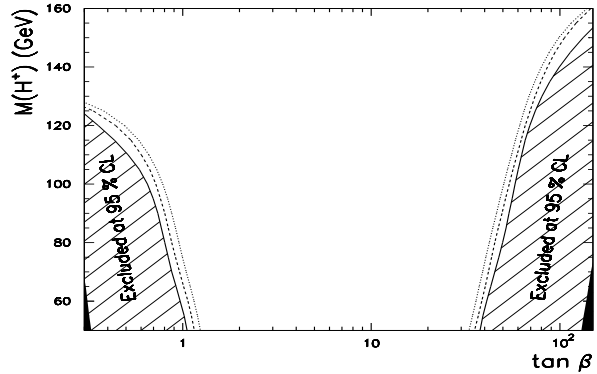, width=\hsize, height=5cm}
\caption{$95\%$ C.L. exclusion contours
  for charged Higgs production for $\mt = 175\gevcc$ and
  $\sigma(t\tbar) = 5.5\pb$ (hatched area, solid lines), $5.0\pb$
  (dashed lines), and $4.5\pb$ (dotted lines).}
\label{fg:chresult}
\end{figure}

The regions in parameter space where our analysis is sensitive are
those where $\BR(t\ra H^+)$ is large.  Those are the regions where
$\mch$ is low, and $\tan\beta$ is either large or small
(see \figref{fg:chmap}).  Once an $H^+$ is produced, it can decay in
several ways.  For large $\tan\beta$, the decay $H^+\ra\tau\nu$ 
dominates, while for small $\tan\beta$, $H^+\ra cs$ is favored.  But
for $\tan\beta$ small and $m_{H^+}$ large, there is an additional
decay mode that becomes important: $H^+\ra t^*b\ra Wbb$.

We have searched for $t\ra H^+$ using a
``disappearance analysis''~\cite{d0chhiggsprl}.  This is based on the
results of 
our cross section measurement in the lepton+jets
channel~\cite{d0xsecprl97}; that is, where one top~quark in a pair decays
via $t\ra bl\nu$ and the other decays via $t\ra bqq$.  In this
channel, we find 30~$\ttbar$ candidates, with an expected background of 
$11\pm 2$ events.  Assuming that the $\ttbar$ cross section has no
contributions from any new physics channels, we can
exclude regions of parameter space in which
$t\tbar$ pairs decay via $H^+$ to final states for which our
event selection has very low acceptance.  In these regions, the
observed excess of signal over background cannot be explained by
$\ttbar$ production.  The result is shown in
\figref{fg:chresult}, for three assumed $\ttbar$ cross
sections.  Note that this analysis is valid only for the interior
of the plot.  LEP excludes $M_{H^+} < 60\gevcc$, while in the other
regions outside the plot, the
perturbative calculations for the Higgs branching ratios
become invalid.  For Run~2, we expect to be able to increase the area
excluded by over a factor of two.

\section{Neural network analysis of $t\tbar\ra e\mu$}

The ``golden'' channel for the observation of $t\tbar$
production is $t\tbar\ra e\mu b\bbar\nu\nu$.  Due to the
presence of two unlike flavor leptons, this channel has very low
background.  But its branching fraction is also a small $2.5\%$.
It is therefore important to maximize the acceptance in this channel.

For our published measurement~\cite{d0xsecprl97}, we required
one electron with $\et > 15\gev$,
one muon with $\pt > 15\gevc$, $\met > 20\gev$, two jets
with $\et > 20\gev$,
$\Delta R(e,\jet) > 0.5$ [$(\eta,\phi)$ space],
$\Delta R(e,\mu) > 0.25$, and
$\htran \equiv \et^e + \sum \et^{\jet} > 120 \gev$.
%
%
For our new analysis~\cite{happythesis}, we
remove the $\htran$ requirement and reduce the jet $\et$ and
$\met$ requirements to $15\gev$.
To regain back\-ground rejection, we turn to a neural network
analysis.

\begin{figure}
\centering
\epsfig{file=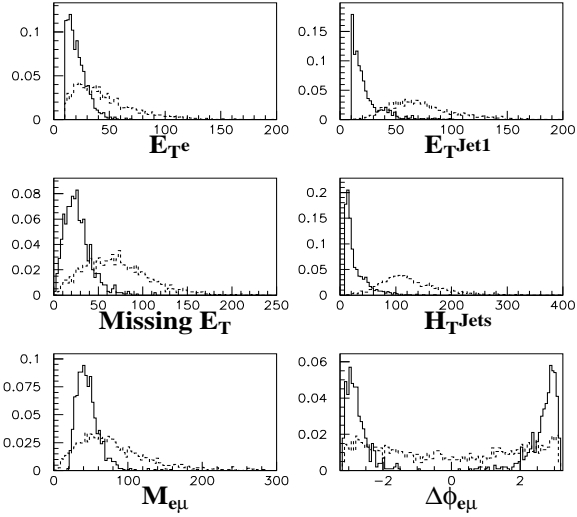, width=\hsize, height=7cm}
\caption{Distributions of input variables to the $\tau\tau$ neural
  network for signal (dashed) and $Z\ra\tau\tau$ background (solid).
  Units are in $\gev$, except for the last plot.}
\label{fg:emu-vars}
\end{figure}

\begin{figure}
\centering
\epsfig{file=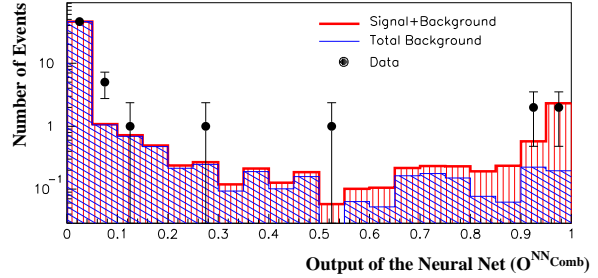, width=\hsize}
\caption{$O_{{\rm NN}}^{{\rm comb}}$ for data, expected background,
  and expected signal plus background.  The
  events below the signal region have a distribution consistent with
  background.}
\label{fg:emu-nncomb-sigbkg}
\end{figure}

There are three major backgrounds: QCD jet production,
$Z\ra \tau\tau \ra e\mu$, and
$WW\ra e\mu$.  A separate network is trained to discriminate the
signal from each of the three backgrounds.  Six variables are used as
inputs to each of the networks, these being $\et^e$, $\et^{\jet 2}$,
$\met$, $H_T^{\jets} \equiv \sum_{\jets} \et^{\jet}$, $M_{e\mu}$, and
$\Delta\phi_{e\mu}$, except for the $\tau\tau$ network, where
$\et^{\jet1}$ replaces $\et^{\jet2}$.  (See \figref{fg:emu-vars}.)
Each of the networks has seven hidden units, and
is trained
on equal numbers of $t\tbar$ signal and background events 
(4000 events for QCD, and 2000 for the other
two).  The outputs $O_{{\rm NN}i}$ are combined using
%
$O_{{\rm NN}}^{{\rm comb}} = 3/\sum_{i=1}^3 1/O_{{\rm NN}i}$.
(See \figref{fg:emu-nncomb-sigbkg}.)
The candidate sample is defined by
$O_{{\rm NN}}^{{\rm comb}} > 0.88$, determined
by maximizing the expected relative sig\-nif\-i\-cance $S/\sigma_B$.
($\sigma_B$ is the uncertainty in the background.)

The results are shown in \tabref{tb:emu-results}.
Compared to the published analysis,
the neural network analysis increases the efficiency
by about $10\%$.  The back\-ground is also slightly lower,
but this is harder to evaluate due its large statistical
uncertainty.

\begin{table}
\centering
\begin{tabular}{ccc}
\hline
    & NN Analysis  & Conventional \\
\hline
    $\epsilon \times \hbox{BR}$ (\%) & 
         $0.402\pm 0.085$  & $0.368\pm 0.078$ \\
    Background              & $0.24\pm 0.15$   & $0.26\pm 0.16$ \\
    Events observed        & $4$              & $3$ \\
    $\sigma(t\tbar)$       & $8.8\pm 5.1\pb$  & $7.1\pm 4.8\pb$ \\
\hline
\end{tabular}
\caption{A comparison of the results of the conventional and neural
  network $t\tbar\ra e\mu$ analyses, for $\mt = 175\gevcc$.  Note that 
  the quoted uncertainties are highly correlated between the two
  analyses.}
\label{tb:emu-results}
\end{table}




\bibliographystyle{reviewsty}
\bibliography{citations}

\end{document}